# Transduction between electrical energy and the heat in a carbon nanotube using a voltage-controlled doping

*T. Gupta[a], I. P. Nevirkovets[a,b], V. Chandrasekhar[a], and S. Shafranjuk[a],**

[a] Physics and Astronomy Dept., Northwestern University, 2145 Sheridan Rd. Evanston IL 60208, USA.
[b] HYPRES, Inc., Elmsford, NY 10523 USA.

**Abstract.**

High electric conductivity $\sigma \geq 2 \times 10^8$ S/m and Seebeck coefficient $\alpha \geq 200$ $\mu$V/K of carbon nanotubes (CNT) make them attractive for a variety of applications. Unfortunately, a high thermal conductivity $\kappa \sim 3000$ W/(m·K) due to the phonon transport limits their capability for transforming energy between the heat and electricity. Here we show that increasing the charge carrier concentrations not only leads to an increase of both $\sigma$ and $\alpha$, but also causes a substantial suppression of $\kappa$ due to intensifying the phonon-electron collisions. A strong transduction effect corresponding to an effective electron temperature change $2\Delta T = 114 \pm 7$ was observed in a CNT device, where the local gate electrodes have controlled the charge doping in the opposite ends. Transduction between the heat and the energy of the electron subsystem corresponds to an impressive figure of merit $ZT_{\text{cold}} = 5.6 \pm 1.7$ and the transduced power density $P_{cooling} \sim 80$ kW/cm$^2$.

## 1. Introduction

Study of the electric and thermal transport on nanoscale is important to better understand the energy transduction mechanisms [1-12] between the heat and electricity inside the low-dimensional materials. [8, 13-16] These have potential applications in many disciplines including nanotechnology [8, 17], remote sensing [19], quantum and conventional digital electronics, sources of clean energy[13], and cooling on nanoscale [16, 20-23]. In many applications, such as sensors and nanoelectronic elements, the conversion between the heat and electrical energy can be localized in nano-size regions. In such systems, the local cooling [16, 18, 20-23] on nanoscale enables to observe and exploit low-temperature phenomena at *ambient temperature*, without any need for bulky and expensive refrigerating equipment.

The method of the local cooling [16, 18, 20-23] uses the low energy electrons drifting through the quantum dot [19, 23], whose effective electron temperature, $T_{el}$, reaches 20-50 K, despite the fact that the ambient temperature, $T_{rm}$, is about 300 K. Recently, the local cooling of individual quantum dots al-lowed the authors[16] to observe the single-electron tunneling and the localized electron energy levels at room temperature [21-23]. The electron thermal excitations were suppressed using the discrete energy levels localized in quantum dots (QD) [16, 22, 24]. During the electron transmission through a discrete energy level $E_C$, the latter works as an energy filter (or as a thermal filter): only the electrons whose energy matches $E_C$ pass through the quantum dot, thereby leading to local cooling of an individual QD. Unfortunately, such "passive" cooling [16, 19, 21-24] has limited capability, since it works for very small systems with a few electrons but fails if the thermal phonons are absorbed. An alternative approach exploits an "active" cooling, e.g., using the solid state thermoelectric transducers [8, 13, 16, 17], which are capable of transforming the heat into electric energy directly. Such materials and devices utilize a dual nature of elementary excitations in conducting materials, which, being the carriers of electric charge, also transmit thermal energy owing to their finite mass.

*Corresponding author. Tel. 847 467-2170
E-mail address: s-shafraniuk@northwestern.edu (S. Shafranjuk).

In this paper, we report observation of transduction between electrical energy and the heat in a semiconducting carbon nanotube (CNT) whose conductivity is controlled by the local gate electrodes. The CNT transducer, schematically shown in Fig. 1, allows for changing the local effective electron

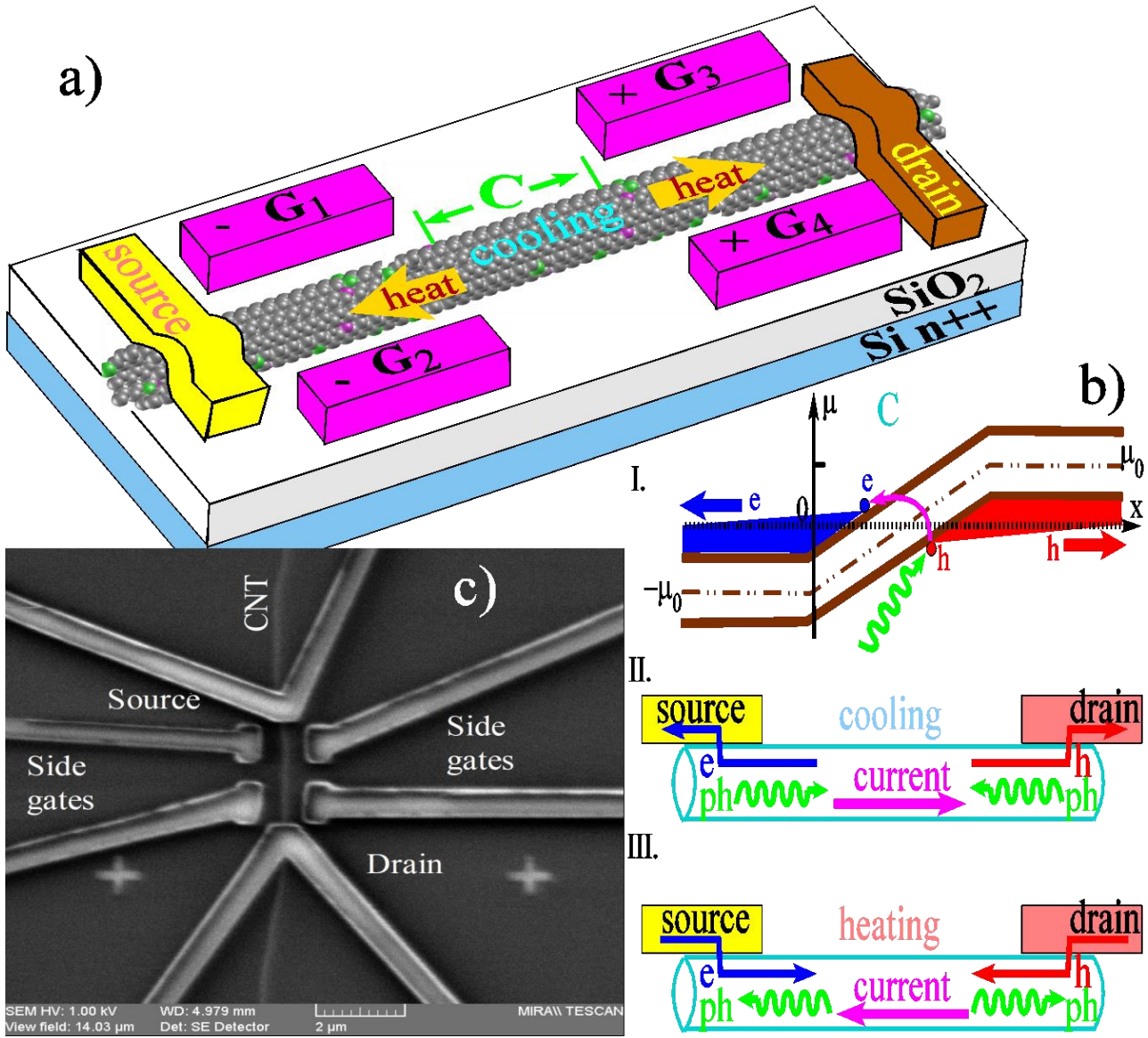

**Figure 1.** (a) Schematic of a transducer made of a single-wall carbon nanotube (CNT) by forming two ambipolar field-effect transistors which are connected electrically in sequence but thermally in parallel. The pairs of side gate electrodes $G_n$ (where n=1..4) are setting the sign and concentration of charge carriers in the CNT. (b) Transduction between the heat and electricity inside the carbon nanotube. I. Energy diagram illustrating the cooling process due to extraction of the electron (blue) and hole (red) excitations from the central region C of the transducer by the transport electric current. Arised deficit of charge carriers in C is compensated by creating of new electrons and holes during the indirect interband transitions (magenta) due to absorption of acoustic phonons (green wave) coming from outside. II. Cooling: Electric current (magenta), flowing along the CNT, pulls out the electron (e) and hole (h) excitations from the central section C of the nanotube, thereby causing cooling. III. Heating: Electric current pushes the electrons and holes from the opposite CNT sides toward the central section, causing heating and infrared emission [18]. c) An SEM image of typical CNT nano-circuit used in the experiments.

temperature $T_{el}(x)$ in the central section C in a wide range. One sets the electron energy profile in the leftmost and rightmost sections of the nanotube, using the local gate electrodes $G_n$, positioned on the

SiO$_2$ substrate close to the CNT on its both sides (see Fig. 1a). Local value of the electrochemical potential $\mu(x)$ inside the CNT is controlled by applying the electric potentials $\varphi_L$ to the gates $G_n$ (see Fig. 1bI). This enables the electrode doping, which introduces either electrons or holes inside the leftmost or rightmost CNT sections, respectively. In particular, by applying a negative electric potential $V_{SG}^{left} < 0$ to $G_{1,2}$ on the left, one creates electrons in the leftmost CNT region, while holes are created in the rightmost section by applying a positive electric potential $V_{SG}^{right}$ to $G_{3,4}$ on the right (see Fig. 1a,b). The central section C of CNT remains undoped, where only electrons and holes in equal numbers are present (see also the Appendix Sec. S1).

A finite bias electric current $I_e$, flowing along the semiconducting carbon nanotube, induces a gradient of charge carrier density along the sample, which leads to a finite temperature difference $\Delta T = T_{hot} - T_{cold}$, where $T_{hot(cold)}$ is the temperature of the hot (cold) part of the sample. Pumping out of electrons and holes from C causes cooling, as seen in Fig. 1b-II. On the contrary, when the direction of $I_e$ is changed to opposite, the electrons and holes are pulled toward the C section, leading to heating, as shown in Fig. 1b-III. Furthermore, strong pumping of electrons and holes from opposite CNT sides toward C causes infrared emission due to recombination [18]. The thermoelectric effect is described as $\Delta V = \alpha \Delta T$ where $\alpha$ is a linear-response, two-terminal property known as Seebeck coefficient.

A combination of basic parameters, determining the efficiency of the thermoelectric process, constitutes the figure of merit $ZT_{cold} = \alpha^2 \sigma T_{cold} / \kappa$, where $\kappa$ is the heat conductivity of the sample. Optimal selection of $\alpha$, $\sigma$, and $\kappa$ represents the major challenge for the successful solution of the thermoelectric energy transformation problem. One can see that $ZT_{cold}$ increases with electric conductivity $\sigma$ and Seebeck coefficient $\alpha$, but decreases with thermal conductivity $\kappa$. Current thermoelectric methods [10-12] still fall short in practical implementations, since the figure of merit of available thermoelectric devices is below the desirable threshold $ZT_c < 4$.

Here we report an experimental observation of thermoelectric cooling of the electron subsystem in the middle region C of CNT (see Fig. 1 and Fig. S1 in Appendix), characterized by the figure of merit $ZT_c = 10 \gg 1$. The CNT transducer can be regarded as a system consisting of two field-effect transistors (FET) operating with the charge carriers of opposite polarity, either electrons or holes. The FETs are connected electrically in sequence but thermally *in parallel* [8]. The setup ensures an appreciable efficiency of the thermoelectric cooling of the electron subsystem. The key element of our thermoelectric nano-circuit is a single-wall semiconducting carbon nanotube situated between the two pairs of the side-gate electrodes $G_n$. The electrodes control the sign and concentration of the charge carriers locally, thereby forming a step-wise electrochemical potential $\mu(x)$ versus coordinate $x$ inside the CNT (see

Fig. 1a,b-I and Fig. S1 in Appendix). The bias voltage $\Delta V$ applied to the nanotube via the source (S) and drain (D) electrodes induces electric current along the CNT. This allows us to change the local effective electron temperature $T_{el}$ in the middle of CNT owing to the transduction effect. The magnitude $\Delta T$ of the temperature change is deduced from the change in the position and width of spectral singularity, which is manifested in the experimental curves of the source-drain electric conductance. Depending on the sign of $\Delta V$, the thermoelectric effect causes either cooling or heating of the electron subsystem inside the CNT, as illustrated by the diagrams shown in Figs. 1b I-III,

Basic parameters $\alpha$, $\sigma$, and $\kappa$, determining the transduction effect in the carbon nanotube, are discussed in Appendix. It is commonly recognized (see, e.g., Refs. [25, 26]) that one may achieve impressive magnitudes of $\alpha$ and $\sigma$ in carbon nanotube junctions. In particular, the reported values [25] of Seebeck coefficient[26] are above 200 μV/K, while $\sigma$ exceeds $2\times 10^8$ S/m. On the one hand, the semiconducting carbon nanotube implemented in our nano-circuit, has appreciable values of $\alpha$ and $\sigma$ [9, 25]. On the other hand, it has a relatively high magnitude of the thermal conductivity $\kappa$, which adversely affects the figure of merit $ZT_{cold}$. The thermal conductivity $\kappa$ is associated with the transport of electrons (e) and phonons (ph) as $\kappa = \kappa_e + \kappa_{ph}$. Typically [3, 6, 8, 9], $\kappa_{ph}/\kappa_e = 10^3 - 10^4$, and thus $\kappa = \kappa_{ph} \gg \kappa_e$. The large value of the thermal conductivity, $\kappa$=100-3000 W/(K·m), poses a well-known disadvantage (see Refs. [3, 6, 8, 9]) of carbon nanotubes when using them as elements of thermoelectric nano-circuits, because it hinders conversion of the heat into the electrical energy, thus limiting the performance of the CNT thermoelectric nano-circuits. Physical mechanisms, reducing efficiency of the thermoelectric CNT nano-circuits - the backflow transport of phonons and the phonon drag - were studied in Refs. [8,13] and are discussed in Secs. 2-4 in Appendix.

Owing to the fact that the phonon contribution to the thermal transport is effectively suppressed due to the phonon-electron collisions (see Sec. 2 in Appendix), the resultant thermal conductance $\Lambda_{\text{CNT}}$ is therefore remarkably low, $\Lambda_{\text{CNT}} = \kappa_{\text{CNT}} \cdot \pi d_{\text{CNT}}^2 / (4L) = 5 \times 10^{-11} - 1.4 \times 10^{-9}$ W/K where $d_{\text{CNT}} \sim 1-3$ nm and $L \sim 3-10$ μm are the CNT diameter and length, respectively. Furthermore, the phonon transport is additionally diminished in the thermoelectric nano-circuit, becoming $\Lambda = \left( \Lambda_{\text{CNT}}^{-1} + 2\Lambda_{\text{Ti/CNT}}^{-1} \right)^{-1} = \Lambda_{\text{Ti/CNT}} / 2 \ll \Lambda_{\text{CNT}}$. This happens because the phonons propagate mostly along the nanotube axis, whereas the phonon mean free path in the electrode doped left and right sections of the CNT becomes considerably shortened. On the contrary, the electrons and holes easily propagate between the CNT and the metallic source and drain electrodes, thereby facilitating the thermoelectric effect (see Figs. 1bII-III). Since the phonons are effectively eliminated from the thermal transport, the overall thermal conductance $\Lambda$ of the whole setup shown in Fig. 1 is by three to four orders of magnitude lower as compared to the known value for the single-wall carbon nanotubes. Therefore, the energy dissipation due to a heat leakage to the substrate can be neglected (see also Sec. S3 in Appendix). This allows one to achieve a considerable increase of the $ZT_c$ value in the setup shown in Figs. 1, S1.

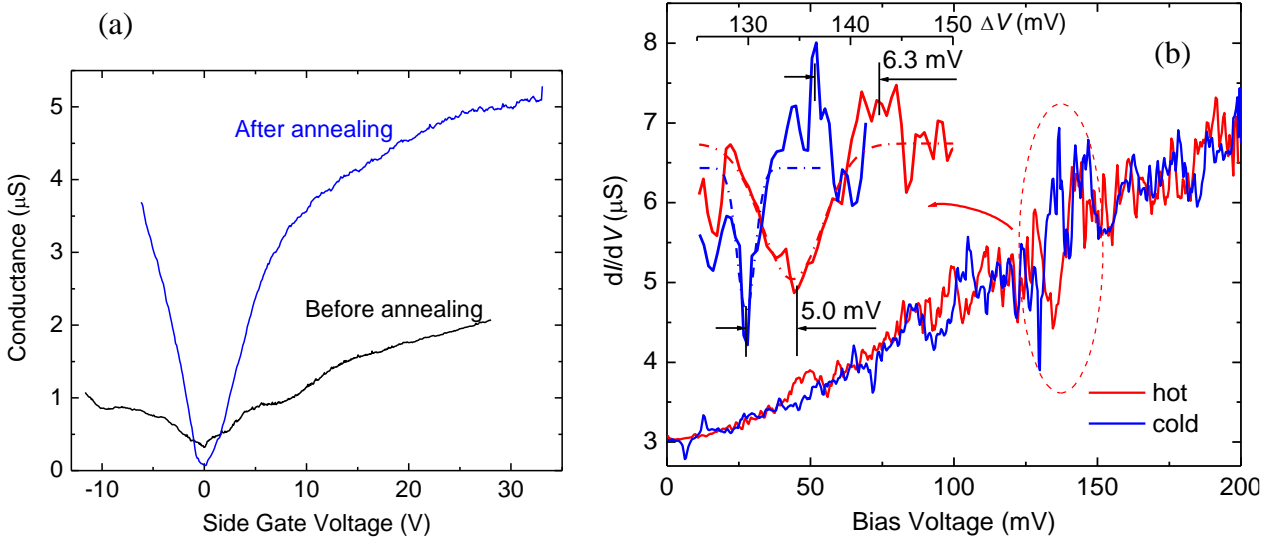

**Figure 2**. (a) Differential conductance of the CNT nano-circuit consisting of two ambipolar field transistors with Ti source and drain electrodes before and after annealing. (b) Determination of the effective electron temperature. Measured differential conductance $G_e(\Delta V)$ of the CNT thermoelectric circuit with Ti electrodes at the bath temperature of 77 K. Blue curve is for the cooling process sketched in Fig. 1bII, whereas the red curve is for the heating process shown in Fig. 1bIII. Localized energy levels are identified as sharp features in the conductivity in the energy interval from 125 to 150 meV (marked by dashed line). A blowup of this region is shown in the inset. From the plot we find that $\Gamma_{\text{cool}} = 1.75 \pm 0.5$ meV and $\Gamma_{\text{hot}} = 5.6 \pm 0.7$ meV giving an estimate of $T_{\text{hot}}$-$T_{\text{cold}}$~$114 \pm 7$ K.

2. **Experimental section**

*2.1. Materials and sample preparation*

The carbon nanotubes were grown in a chemical vapor deposition (CVD) chamber using an optimized catalyst and annealing process. The catalyst consists of $Fe(NO_3)_3 \cdot 6H_2O$ (0.2 g) and Alumina nano-particles (5 nm, 0.4 g) in 10 ml of deionized water. The substrates, p-doped silicon with a 300 nm or 1 μm thermal oxide layer, are then dipped in the solution such that only a thin layer of solution is left on the substrate edge. Traditionally, the catalyst solution is spun onto the substrate; we have found that while this yields a large number of tubes they are densely packed and curved, thus being unsuitable for nano-fabrication. The dipping method, however, yields a low to moderate density of tubes, depending strongly on the concentration of $FeNO_3$ in the solution. We position the substrate inside the CVD chamber in such a way that the catalyst dipped edge is perpendicular to the flow of gases. This is done so as to get the tubes to fall in the catalyst-free region of the substrate.

CNT grown at higher temperature are usually linear and their size can go up to a few centimeters in length. Hence, they are ideal for fabricating of nano-circuits. During the early stages of fabrication, our focus was to find nanotubes which had ambipolar field-effect transistor (FET) behavior and low interface resistance. Summary of the CNT growth conditions used in the experiments is given in Table 1.

**Table 1.** The experimental parameters used to grow nanotubes and the variations produced by the different growth methods.

| Growth Temp. (°C) | $CH_4$ flow (sccm) | $H_2$ flow (sccm) | Argon flow (sccm) | Comment |
|---|---|---|---|---|
| 900 | 900 | 140 | 100 | Usual method, long tubes |
| 800 | 450 | 70 | 50 | Growth is similar to 900 °C condition |
| 700 | 450 | 70 | 50 | Shorter tubes and low density |
| 1000 | 100 | 350 | 0 | cm-long straight tubes |

*2.2. Apparatus*

The nanotubes were first characterized using both AFM and SEM techniques. From these measurements, and from electric characterization, we determined that the nano-tubes grown at 900°C were semiconductor-like. The CNT thermoelectric nano-circuits were fabricated using conventional electron beam lithography on a single CNT, thin-film metal deposition, and lift-off. SEM images of a typical fabricated structure is shown in Fig. 1c. The nano-circuit consists of the central wire (CNT), and the side gates fabricated parallel to the CNT, and spaced nominally from 300 to 400 nm from the CNT. Finally, along the CNT itself, a conventional four-probe structure is fabricated in order to both characterize the tube and drive the thermoelectric circuit. After all of the structures are fabricated and connected to larger bonding pads, the area surrounding the tube, which is protected via resist, is cleaned of any stray tubes using oxygen etching. This is done to ensure that the side gates are not shorted to the device.

The ambipolar CNT field effect transistor behavior had been obtained using the titanium contacts with the thickness of about 50 nm. Initially, the devices had too high interface resistance; Annealing such devices at 270 °C for 20 minutes reduces the contact resistance to about 20 kΩ, A typical FET behavior for a CNT device with Ti contacts before and after annealing is shown in Fig. 2(a).

## 3. Results and discussion

### 3.1. Localized electron spectral singularity

Central result of this work is presented in Fig. 2(b) where the differential conductance d$I$/d$V(\Delta V)$ ≡ $G_e(\Delta V)$ in units of microSiemens is plotted against the bias voltage across the CNT. The dependence displays hysteresis for different directions of sweeping the bias voltage. This type of hysteresis is not sensitive to annealing but instead strongly depends on the difference of the side gate potentials. We attribute the hysteretic behavior to manifestation of the energy levels[27] localized in the central part C of the nano-circuit (cf. Fig. 1a,c, 1S). Their position and width depends on the difference of the side gate potentials $V_{SG}^{right} - V_{SG}^{left}$, as is evidenced by respective features in the $G_e(\Delta V)$ curves (cf. Fig. 2(b). The features depend on a change $\Delta T$ of local temperature $T$ versus direction of the electron current (cf. Figs. 1bI and 1bII). With an appropriate calibration, it allows for mapping [28] the level shift $\Delta E$ and broadening $\Gamma$ to the temperature $T$ of the active region. At $V_{SG}^{right}$ = 25 V and $V_{SG}^{left}$ = -3.5 V for the curves in Fig. 2(b), the individual FET conductance was about 10.5 μS. Individual FETs are 2 μm each and the central part is 0.8 μm. The entire length of the device was around 5 μm. We used this circumstance to identify the energy levels $E_C$ localized in the active region C, as is shown in the inset of Fig. 2(b).

### 3.2. Deducing of the intrinsic electron temperature

The obtained experimental data allowed us to determine the local intrinsic temperature change of the middle CNT region as follows. Because the local intrinsic temperature depends on the coordinate $x$ along the CNT axis, and the local resistance $R_{CNT}(T,x)$ of the CNT is temperature-dependent, this allows for identifying the energy level $E_C$ localized in the central active region C from sharp features in the differential source-drain conductance curves $G_e(\Delta V)$. The local effective electron temperature

$T_{el}(x)$ is inferred [28] by measuring the level width $\Gamma_{cold(hot)}$ and the shift $\Delta E_0$ of the position of an energy level $E_C$ localized in the active central (C) region of CNT.

The temperature dependence $\Gamma_0(T)$ is derived [28] by fitting the sharp features with the Gauss bell curves as was suggested earlier in Ref. [29]. Comparing the $G_e(\Delta V)$ curves, measured at different $\Delta V$ corresponding to a variety of $T_{el}(x)$ in the region C, we identify the localized levels $E_C$, whose position and width are changed versus $\Delta V$. The shifts $\Delta_{1,2}$ of the level positions occur owing to the electric resistance change of the active region C during the thermoelectric heating and cooling cycles. The sharp features in the $G_e(\Delta V)$ curve, corresponding to the energy levels $E_C$ localized in the active region, are shown in the inset of Fig. 2(b). Let us consider the two minima in the $G_e(\Delta V)$ curve, visible in Fig. 2(b). One can notice that the width $\Gamma_{hot(cold)}$ of the mentioned minima is significantly different for the red (hot) and blue (cold) curves, i.e., $\Gamma_{hot} \gg \Gamma_{cold}$. By fitting the experimental data with the Gauss bell curves (dash-dotted curves in the inset of Fig. 2(b) we find $\Gamma_{hot} = 5.6 \pm 0.7$ meV and $\Gamma_{cool} = 1.75 \pm 0.5$ meV where the corresponding fitting error $\Delta\Gamma$ is determined by the number of experimental points used for averaging. The large magnitude of $\Delta\Gamma$ is caused by a considerable noise level during the measurement. The mapping of the level width $\Gamma_{hot(cold)}$ to the intrinsic temperature $T_{hot(cold)}$ of the active region is accomplished by determining the dependence of the level width $\Gamma_0(T_{bath})$ versus the bath temperature $T_{bath}$ (see Ref. [28]) from the $G_e(\Delta V)$ curves measured when setting $V_{SG}$=0. The steady-state temperature dependence $\Gamma_0(T_{bath})$ is then used for calibrating the local temperature $T_C$ at finite $V_{SG}$ [28, 29].

*3.3. The deduced figure of merit*

Our experimental results suggest that the electric current along the nanotube induces an impressive change of local temperature $2\Delta T = 114 \pm 7$ K inside the central CNT section. Depending on the direction of the source-drain current, the temperature either increases from the liquid nitrogen temperature $T = 77$ K up to $T_{hot} = 134 \pm 8$ K, or decreases from $T = 77$ K down to about $T_{cold} \simeq 20 \pm 6$ K, thus evidencing a strong thermoelectric effect. We determine the dimensionless figure of merit $ZT_{cold}^2$ using a condition that the maximum temperature change $\Delta T_{max}$ is defined by $\Delta T_{max} = G_e \mathbf{s}^2 T_{cold}^2 / (2\Lambda) = ZT_{cold}^2 / 2$, giving an impressive value $ZT_{cold} = 2\Delta T_{max} / T_{cold} = 5.6 \pm 1.7$ for our nano-circuit.

The figure of merit $ZT_{cold}$ can be further improved by increasing the source-drain bias voltage, because this allows one to pull out more electrons from the C section toward the left CNT end and holes toward the right CNT end due to increasing the bias electric current. In this experiment, we used a maximum voltage of 200 mV, where the effect of cooling is expected to be higher than at $\Delta V$~130 mV (cf. Fig. 2(b)). Although we have not observed any spectral singularities at a higher $\Delta V \sim 200$ mV, which can be used for determination of $\Delta T$, we can extrapolate the values deduced for ~130 mV (i.e., using $2\Delta T = 114 \pm 7$ and $ZT_{cold} = 5.6 \pm 1.7$ as the reference points). Then, applying the energy balance condition derived in S4 of Appendix, we obtain the values of $\Delta T = 64 \pm 8$ K and $ZT_{cold} = 10 \pm 2.8$. Estimated cooling power density is $P_{cooling}$ ~ 80 kW/cm² for our CNT transducer where $R_{contact}$ ~ 100 kΩ with the CNT/Ti contacts. The overall transducing power can be increased additionally by decreasing $R_{contact}$, increasing $\Delta V$ and the local gate voltages $V_{SG}^{right} - V_{SG}^{left}$, and also by scaling up to large CNT networks and arrays.

## 4. Conclusion

In conclusion, we studied the transduction of the heat and electrical energy inside a carbon nanotube. In contrast to conventional semiconducting transducers, where increasing the carrier concentration causes an increase of $\sigma$ but reduces $\alpha$ and increases $\kappa$, the CNT shows a remarkably different behavior. When the charge carrier concentration is increased, $\sigma$ and $\alpha$ both grow, while $\kappa$ decreases. This results in appreciable *ZT* and densities of transduced power $P_{cooling}$~80 kW/cm²; this magnitude can be increased further using the voltage-controlled spectral singularities and filtering the electric/heat currents, as suggested in Refs. [8] [13].

**Acknowledgments**

We wish to thank K. Novoselov, M. S. Dresselhaus, and G. Chen for fruitful discussions. We are also grateful D. Dikin, S. Davis and S. Mayle for technical help. We acknowledge support from the AFOSR grant FA9550-11-1-0311.

**Appendix. Supplementary information**

Supplementary information including the theoretical model and detailed estimation of experimental parameters can be found at

**References**

1. Bell, L. E. *Science* **2008,** 321, (5895), 1457-1461.
2. DiSalvo, F. J. *Science* **1999,** 285, (5428), 703-706.
3. Berber, S.; Kwon, Y. K.; Tomanek, D. *Phys Rev Lett* **2000,** 84, (20), 4613-4616.
4. Dubi, Y.; Di Ventra, M. *Reviews of Modern Physics* **2011,** 83, (1), 131-155.
5. Finch, C. M.; Garcia-Suarez, V. M.; Lambert, C. J. *Phys Rev B* **2009,** 79, (3).
6. Pop, E.; Mann, D. A.; Goodson, K. E.; Dai, H. J. *J Appl Phys* **2007,** 101, (9).
7. Reddy, P.; Jang, S. Y.; Segalman, R. A.; Majumdar, A. *Science* **2007,** 315, (5818), 1568-1571.
8. Shafranjuk, S. E. *Epl-Europhys Lett* **2009,** 87, (5).
9. Small, J. P.; Perez, K. M.; Kim, P. *Phys Rev Lett* **2003,** 91, (25).
10. Heremans, J. P.; Dresselhaus, M. S.; Bell, L. E.; Morelli, D. T. *Nat Nanotechnol* **2013,** 8, (7), 471-473.
11. Zhao, L. D.; Lo, S. H.; Zhang, Y. S.; Sun, H.; Tan, G. J.; Uher, C.; Wolverton, C.; Dravid, V. P.; Kanatzidis, M. G. *Nature* **2014,** 508, (7496), 373-+.
12. Snyder, G. J.; Toberer, E. S. *Nat Mater* **2008,** 7, (2), 105-114.
13. Shafranjuk, S. E. *European Physical Journal B* **2014,** 87, (4).
14. Shafranjuk, S. E. *Epl-Europhys Lett* **2015,** 109, (6).
15. Mayle, S.; Gupta, T.; Davis, S.; Chandrasekhar, V.; Shafraniuk, S. *J Appl Phys* **2015,** 117, (19).
16. Shafraniuk, S., *Graphene: Fundamentals, Devices and Applications*. Pan Stanford: 2015; p 634.
17. Rowe, D. M., *Thermoelectrics handbook : macro to nano*. CRC/Taylor & Francis: Boca Raton, 2006.
18. Avouris, P.; Freitag, M.; Perebeinos, V. *Nature Photonics* **2008,** 2, (6), 341-350.
19. Rinzan, M.; Jenkins, G.; Drew, H. D.; Shafranjuk, S.; Barbara, P. *Nano Letters* **2012,** 12, (6), 3097-3100.
20. Takahashi, Y.; Nagase, M.; Namatsu, H.; Kurihara, K.; Iwdate, K.; Nakajima, K.; Horiguchi, S.; Murase, K.; Tabe, M. *Electronics Letters* **1995,** 31, (2), 136-137.
21. Xiong, Z. W.; Wang, X. M.; Yan, D. W.; Wu, W. D.; Peng, L. P.; Li, W. H.; Zhao, Y.; Wang, X. M.; An, X. Y.; Xiao, T. T.; Zhan, Z. Q.; Wang, Z.; Chen, X. R. *Nanoscale* **2014,** 6, (22), 13876-13881.
22. Zheng, H. S.; Asbahi, M.; Mukherjee, S.; Mathai, C. J.; Gangopadhyay, K.; Yang, J. K. W.; Gangopadhyay, S. *Nanotechnology* **2015,** 26, (35).
23. Bhadrachalam, P.; Subramanian, R.; Ray, V.; Ma, L. C.; Wang, W. C.; Kim, J.; Cho, K.; Koh, S. J. *Nature Communications* **2014,** 5.
24. Shafraniuk, S. E., Nanosensors of External Fields. In *Encyclopedia of Nanoscience and*

*Nanotechnology*, Nalwa, H. S., Ed. American Scientific Publishers: 2011; Vol. 18, pp 413-454.

25.     Nakai, Y.; Honda, K.; Yanagi, K.; Kataura, H.; Kato, T.; Yamamoto, T.; Maniwa, Y. *Applied Physics Express* **2014,** 7, (2).

26.     Zuev, Y. M.; Chang, W.; Kim, P. *Phys Rev Lett* **2009,** 102, (9).

27.     Gerritsen, J. W.; Shafranjuk, S. E.; Boon, E. J. G.; Schmid, G.; vanKempen, H. *Europhys Lett* **1996,** 33, (4), 279-284.

28.     Y. Yang, G. F., S. E. Shafranjuk, T. M. Klapwijk, B. K. Cooper, R. M. Lewis, C. J. Lobb, and P. Barbara. *Nano Letters* **2015**.

29.     Yang, Y. F.; Fedorov, G.; Zhang, J.; Tselev, A.; Shafranjuk, S.; Barbara, P. *Superconductor Science & Technology* **2012,** 25, (12).

**Appendix: Parameters and theoretical model of the CNT transducer**

**S1. Peltier cooler efficiency and the phonon contribution to the thermal flux**

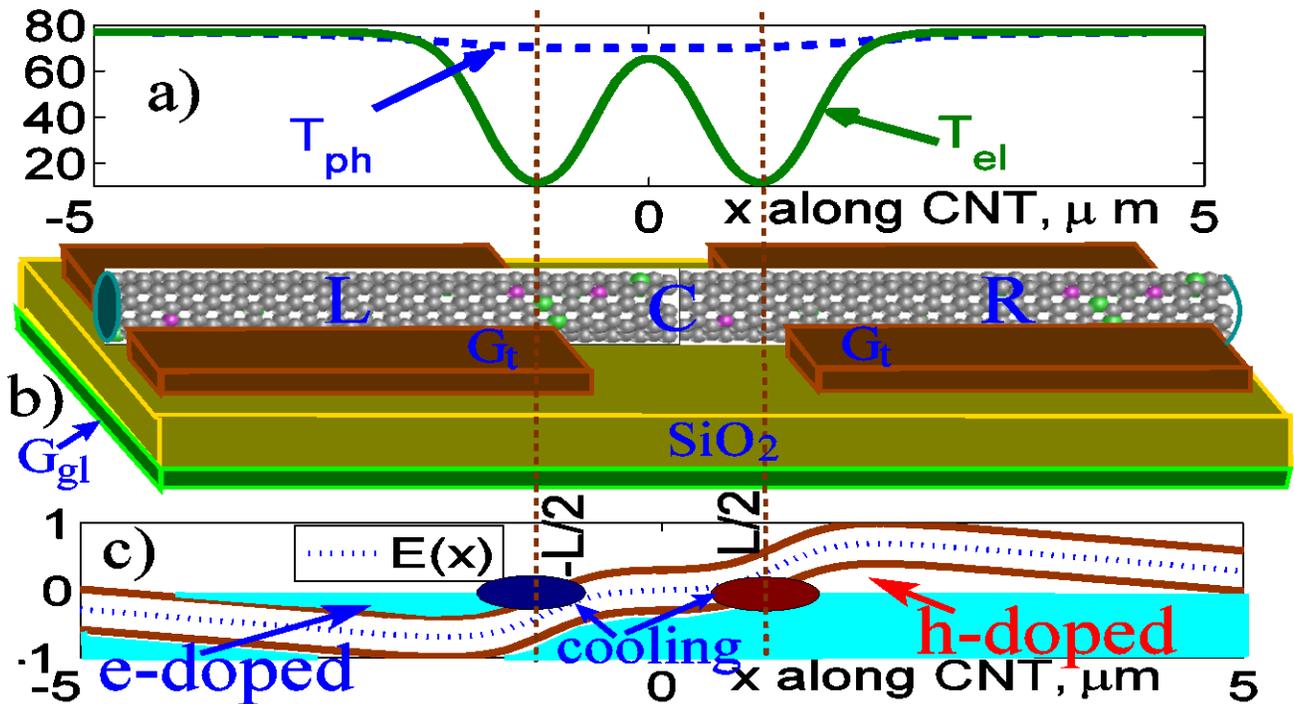

**Figure S1 | Schematics of the Peltier cooling in CNT.** (a) Spatial distribution of the effective temperature of electrons $T_{el}(x)$ and phonons $T_{ph}(x)$ along the CNT for the bias voltage $\Delta V = 135$ V (see calculation details in Sec. A4). The electron temperature $T_{el}(x)$ sharply drops at $-L/2 < x < L/2$ (i.e., in the active region C) owing to extraction of the electron and hole charge carriers from the C region toward the L and R sections by the flowing electric current. (b) The Peltier cooler formed by a CNT, resting on the dielectric SiO$_2$ substrate. $G_{gl}$ is the global gate establishing the Fermi energy value, $G_t$ are the local gate electrodes. (c) The voltage biased CNT structure with the electrode doping of the L and R sections. The electron spectrum becomes dependent on the coordinate $x$ along the CNT, since its profile is set by applying the electric potentials $\varphi_L$ to the $G_t$ electrodes. The cooling takes place at the edges of the active region enclosed between the brown dash lines, in the areas denoted by the blue and red ellipses indicating where the electrons and holes respectively are created.

The idea of the carbon nanotube Peltier cooler is shown in Fig. S1. The thermoelectric cooling proceeds as follows. An electric current, flowing along the CNT from left to right, forces the electrons to the drift toward the L region, whereas the holes are drifting in the opposite direction, i.e., toward the R region. In this way, one extracts the charge carriers from the C region toward the L and R regions, thereby creating a deficit of the hole and electron excitations in the vicinity of C. The deficiency of the electron and hole populations causes a decrease of the local effective temperature $T_{el}(x)$ of electrons at $-L/2 < x < L/2$ (i.e., in the active region C) to a level far below the ambient temperature $T$ (see the corresponding calculations in Sec. S4 below). Simultaneously, in the course of the electron-phonon collisions, considered in Sec. S2, the lowered concentration of the charge carriers tends to re-establish itself back to its equilibrium value, due to creating of new electron-hole pairs caused by absorption of the thermal phonons, thereby transferring energy from the phonon subsystem to the electron subsystem. The energy transfer process is accomplished in the course of the phonon-electron collisions as described in Sec. S4 below. Therefore, the local effective temperature $T_{ph}(x)$ of the phonon subsystem at the ends of C region also is lowered as compared to the ambient temperature $T$, as seen in Fig. S1a, where $x$ is the coordinate along the nanotube. Thus, inside the central region C, the electrons and holes are characterized by lower effective temperature $T_{el}$, provided that $T > T_{ph} > T_{el}$, as shown in Fig. S1a. The precise relationship between $T$, $T_{ph}$ and $T_{el}$ depends on the amount of thermal energy, which has been transferred owing to backflow of the phonons from the L and R sections into the C section (see Fig. S2). The exact value of the transferred energy is determined by the phonon heat conductance $\Lambda_{ph}$ of the charge-doped L and R sections. Because the effective temperatures $T_{el}$ and $T_{ph}$ of the phonon and electron subsystems differ from each other, this initializes the heat energy transfer inside the C section from the phonon subsystem to the electron subsystem.

The dimensionless figure of merit for the cooling process is defined as

$$ZT_{cold} = \frac{S^2 G_e}{\Lambda} T_{cold}, \tag{S1}$$

where $T_{cold}$ is the temperature of the cold region. The material parameters in Eq. (S1) involve Seebeck coefficient $S$, the electric conductance $G_e$ and thermal conductance $\Lambda$; the parameters determine the efficiency of the Peltier cooling [8, 13, 15, 16, 30-34] in the system shown in Fig. S1. From Eq. (S1) one can see that $ZT_{cold}$ increases with $S$ and $G_e$, and decreases with $\Lambda$. On the one hand, both $S$ and $G_e$ are roughly proportional to the electron scattering time $\tau_e$ that depends on the CNT purity and electron-phonon collisions [35-37]. On the other hand, in the absence of doping, the thermal conductance $\Lambda$ is determined mostly by the scattering time $\tau_p$ of phonons on other phonons and also depends on the interface roughness [38-44]. In the course of propagation of the electron and hole excitations along the nanotube, they transfer part of their energy at the expense of the electron-phonon collisions to the phonon subsystem. A fraction of so excited phonons propagates along the nanotube, carrying the obtained energy away from the C section towards the nanotube ends. In the course of the reverse phonon-electron collisions [44], a fraction of the obtained energy is returned back to the electrons and holes, which eventually carry it towards the metal S and D electrodes, as denoted by the blue and pink arrows in Fig. S2.

However, there are other phonons, which do not propagate along the nanotube but escape to the substrate [45-47]. Those phonons are also responsible for the energy dissipation, since after escaping to the substrate, they disappear in the ambient environment and cannot return back to the nanotube. The relevant energy losses can be described as an effective reduction of the electron scattering time $\tau_e$, resulting in diminishing of $S$ and $G_e$, and therefore, in a reduction of the value of $ZT_{cold}$. We evaluate the substrate effect in Sec. S3. It is instructive to estimate the ratio of the phonons escaping into the substrate to the phonons reaching the metal electrodes. Therefore, below we discuss the phonon scattering processes determining their mean free path.

The phonon mean free path $l_{p-p}$ due to phonon-phonon collisions for three-phonon umklapp processes, assuming $\hbar\omega/k_BT \gg 1$, is estimated [48] at the liquid nitrogen temperature $T = 77$ K as

$$l_{p-p} = \frac{c_m A}{\omega^2 T} \simeq 30 \ \mu\text{m}, \tag{S2}$$

where $A = 3.35 \times 10^{23}$ m·K/s$^{-2}$ is the coupling constant in graphene, $c_m = 0.65$ is the parameter of the CNT curvature, and a typical frequency of thermally excited phonons is $\omega = k_BT/\hbar \simeq 10^{13}$ s$^{-1}$. The corresponding phonon-phonon scattering rate is evaluated as

$$\Gamma_{p-p} = \hbar \frac{s}{l_{p-p}} \simeq 0.5 \ \mu\text{eV}. \tag{S3}$$

The above Eq. (S2) suggests that the electron mean free path $l_{p-p}$ due to three-phonon umklapp processes is much longer than the dimensions of the L, R, and C sections, and thus such scattering can be disregarded.

**S2. Electron-restricted phonon scattering in the gated CNT sections**

In the process of Peltier cooling, the phonon flux comes from the outside environment toward the active region, i.e., in the opposite direction to the propagating electrons and holes (see Fig. S2). This part of the external heat flux is diminished owing to the effect of phonon-electron collisions taking place in the charge doped CNT sections. In an undoped CNT, in the absence of charge carriers, the phonon-electron scattering time formally is set as $\tau_{p-e} = \infty$. However, by applying the gate voltage, one introduces the charge carriers - electrons and holes - into the nanotube. Then, on the one hand, $\tau_{p-e}$ becomes finite outside the active region, since the charge carriers serve as scattering centers for the propagating phonons. On the other hand, in the undoped middle (active) region, the phonon scattering is determined by the phonon-phonon collisions, while the phonon-electron scattering is negligible. Therefore, in the middle C region, where the carriers of electric charge are absent, the phonon propagation is almost ballistic. On the contrary, outside the middle C region, where the gate voltages applied to the $G_t$ electrodes induce the finite concentration of charge carriers in the L and R regions (see Figs. S1, S2), the phonons scatter on the electrons and holes [44], thereby transmitting their energy to the charge carriers. The electron concentration $n_e(x)$ versus coordinate $x$ is assumed to have a step-wise form. To find $n_e(x)$ more accurately, one can use, e.g., the approach [49, 50]. Owing to the phonon flux from outside the cooling device, the charge carriers, coming from the middle active region, are confined to the metal S and D electrodes. For an estimation of the electron-phonon scattering rate one can use the well-known expression [51]

$$\frac{1}{\tau_{p-e}} = \eta_{ps} \frac{n_e C_0^2 \omega}{\rho s^2 k_B T} \sqrt{\frac{\pi m s^2}{2 k_B T}} \exp\left(-\frac{m s^2}{2 k_B T}\right), \tag{S4}$$

where $s$ is the acoustic phonon velocity, $n_e$ is the concentration of conduction electrons, $C_0^2$ is the deformation potential, $\rho$ is the mass density, and $m$ is the electron effective mass, typically $m = 0.1-0.5 m_e$ [35-37]. The above formula (S4) is modified owing to pseudospin conservation effects [52], whose contribution is accounted for by introducing an additional factor $\eta_{ps} = 0.3$. Here we assume that the phonon confinement does not strongly affect the phonon-electron scattering rates. We use that [35-37] $\rho = 7.6 \times 10^{-8}$ g/cm$^2$, $s = 2 \times 10^6$ cm/s, $C_0 = 19$ eV, $n_e = 3.671 \times 10^9$ m$^{-1}$, which gives $\tau_{p\text{-}e}^{-1} = 1.5 \times 10^{11}$ s$^{-1}$. The effect of the global gate (which is formed at the bottom of the dielectric substrate) on $n_e$ can be evaluated using the formula for an electric potential difference $V_G$ between the conducting wire with diameter $d$ and the horizontal plane separated by a distance $h$

$$V_G = \frac{\lambda}{2\pi\varepsilon\varepsilon_0} \ln\frac{4h}{d} \tag{S5}$$

where $\lambda$ is electric charge per unit length. The above formula (S5) gives

$$\lambda = \frac{2\pi\varepsilon\varepsilon_0}{\ln(4h/d)} V_G \tag{S6}$$

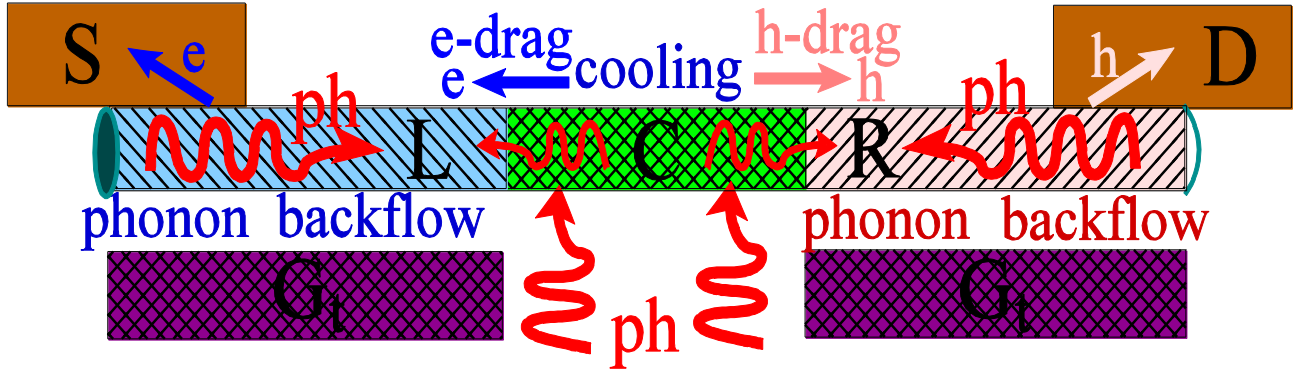

**Figure S2 | Phonon transport inside the CNT Peltier cooler.** The phonons are pushed out of the central cooled C region C due to collisions with the electrons drifting toward the L region of the CNT and with the holes drifting toward the R region. The corresponding microscopic mechanism is the e- and h-drag, originating from electron-phonon collisions. The electrons and holes, drifting in opposite directions under influence of the electric current, collide with phonons, thereby creating the force, which pushes the phonons toward the nanotube ends. On the contrary, since the phonon density at the nanotube ends is higher than in the central region, the backflow of phonons happens from the L and R regions toward the central C region. Owing to a high concentration of electrons in L region and holes in R region of CNT, the phonon-electron scattering causes a significant shortening of the phonon mean free path in these regions. Because the charge carrier concentration in the undoped C region is relatively low, the phonon mean free path there is large.

Then we find the electron density per unit length $n_e$ induced by the gate voltage $V_G$ in the CNT as

$$n_e = \frac{2\pi\varepsilon\varepsilon_0}{e\ln(4h/D)} V_G = 3.7 \times 10^9 \, \frac{1}{\text{m}}. \tag{S7}$$

We used that the dielectric constant of the SiO$_2$ substrate is $\varepsilon = 3.9$ and the gate voltage $V_G = 10$ V. Finally, for the electron density $n_e = 3.7 \times 10^9$ m$^{-1}$, using the CNT section length $l_e = 1-10$ μm, we obtain $\tau_{p\text{-}e}^{-1} = 1.5 \times 10^{11}$ s$^{-1}$ and $\Gamma_{p\text{-}e} = 40$ μeV. More generally, using the phonon-electron scattering rate $\Gamma_{p-e} = \hbar/\tau_{p\text{-}e} = 0.05 - 0.5$ meV, we estimate the phonon mean free path $l_{p\text{-}e}$ due to the phonon-electron collisions as $l_{p\text{-}e} = s \cdot \tau_{p-e} = 30 - 300$ nm, where we used that $s = 2.1 \times 10^4$ m/s for LA phonons. In a similar way, one describes effect of the side gates. In principle, by applying higher gate voltages, which are close to the breakdown voltage of the dielectric substrate (for the SiO$_2$ substrate by thickness $d_{SiO2} = 300$ nm the maximum gate voltage achieves ~1 kV), one can achieve much higher electron concentrations in the gated region of CNT, $n_e = 2 \times 10^{13}$ m$^{-1}$, which gives a very short phonon mean free path $l_{p\text{-}e} \simeq 5-50$ nm owing to the phonon-electron collisions.

This allows for estimating the phonon part of the thermal conductivity restricted by the electron scattering as

$$\kappa_{ph} = \frac{\rho_{2D}}{d_G} c \frac{s \cdot l_{ph}}{3} = 5 - 50 \frac{\text{W}}{\text{m} \cdot \text{K}}$$

where we used the mass density of graphene $\rho_{2D} = 7.6\times10^{-8}\,\text{g/cm}^2$, the graphene layer thickness $d_G = 0.34$ nm, the electron-restricted phonon mean free path $l_{p-e} = 30-300$ nm, and the specific heat of graphene $c_G = 10\,\text{mJ/g}\cdot\text{K}$. The above estimate suggests that inside the L and R sections (i. e. outside the C region), where the local gate voltages are applied, the phonons transfer their energy to the electrons and holes owing to a much slower phonon group velocity ($s/v_F = 10^{-2}$). Furthermore, in the doped L and R regions, the phonon transport is effectively blocked owing to intensive phonon-electron collisions. In accordance with the Fourier law, the heat flux between the CNT and the substrate vanishes because the effective temperature of phonons $T_{ph}$ is equal to the temperature of the substrate. Due to the aforementioned reasons, the energy transfer from the carbon nanotube to the metal electrodes outside the active region occurs solely at the expense of the electron transport.

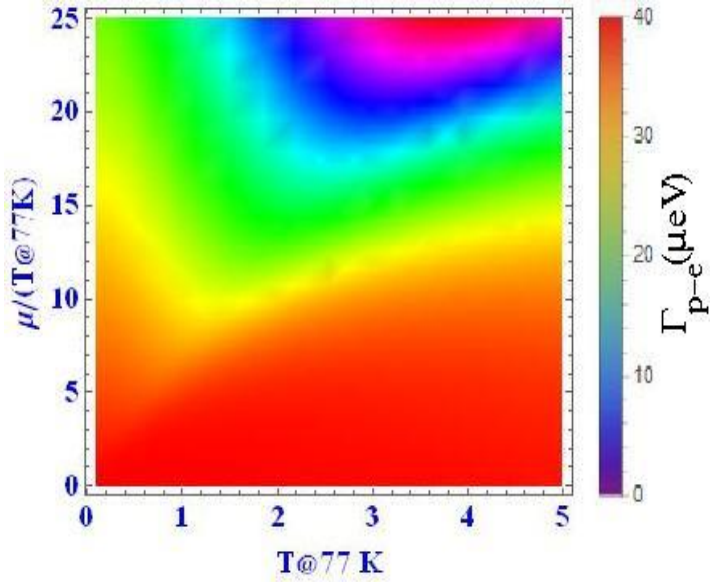

**Figure S3 | The phonon-electron scattering rate** $\Gamma_{p\text{-}e}$ (μeV) versus effective electron temperature $T_{el}$ in units of the liquid nitrogen temperate nitrogen temperature 77 K. One can see that the ratio $\Gamma_{p\text{-}e}/\Gamma_{p\text{-}p} \approx 10^2$, thereby suggesting that owing to the phonon-electron scattering, the phonon mean free path is shortened by two orders of magnitude, as compared to the phonon-phonon scattering mechanisms.

According to Eq. (S1), other important characteristics determining the dimensionless figure of merit $ZT_{cold}$ are the electric conductance $G_e$ and Seebeck coefficient $S$. For the carbon nanotube structure sketched in Fig. S1, both $G_e$ and $S$ are determined by the number of conducting channels $N_{ch}$ inside the CNT and by the transparency $\bar{T}$ of the contact between the CNT and metal electrodes, whereas the other mechanisms like the electron-phonon scattering and escape to the substrate can be neglected. This gives the conductivity

$$\sigma = \frac{L}{A}\frac{2e^2}{h}N_{ch}\bar{T} = 2\times 10^8 \frac{\text{S}}{\text{m}} \tag{S8}$$

where we used for the gated CNT section $L = 5$ μm, $\bar{T} = 0.4$, and the CNT cross section area $A = \pi d_{CNT}^2/4 = 3\,\text{nm}^2$ (where the CNT diameter is $d_{CNT} = 2$ nm). According to Ref. [25], Seebeck coefficient for an undoped semiconducting CNT achieves $S \simeq 0.2\times 10^{-3}$ V/K, which in our setup shown in Fig. S1 is improved further by using the electrode doping, yielding $S$ up to $0.5 \times 10^{-3}$ V/K. Using the above parameters, one estimates the figure of merit as high as $ZT_{cold} = 10 - 100$.

### S3. Thermal flux between the CNT and the SiO$_2$ substrate

Part of the thermal flux, carried by the phonons, goes from the CNT into the SiO$_2$ substrate [45-47], while the other part is directed along the CNT. For the latter fraction, the Fourier law gives

$$q_\| = -\kappa_\| \frac{dT}{dx} = -\kappa_\| \frac{T_h - T_c}{L/2} = -4.9 \times 10^{11} \frac{W}{m^2}, \quad (S9)$$

where $q_\|$ is the heat flux density, $\kappa_\| = 3600 \, W/(m \cdot K)$ is the heat conductivity of CNT, $\nabla T$ is the temperature gradient, and we used $T_h - T_c = 70 \, K$ and $L = 1 \, \mu m$. The corresponding heat flux along the nanotube with diameter $d_{CNT} = 3 \, nm$ is $Q_x = \pi d_{CNT}^2 \cdot q_x = 1.4 \times 10^{-5} \, W$. The boundary thermal conductance, according to Ref. [53] is $K = 11$ nW/K, which for the temperature difference $T_h - T_c = 70$ K gives the heat flux $Q_y = -K(T_h - T_c) = 1 \, \mu W$. Thus, the fraction of energy, which the acoustic phonons carry from the CNT down to the substrate is

$$\frac{Q_\perp}{Q_\|} \simeq 0.1. \quad (S10)$$

Furthermore, using the known value of thermal conductance $\kappa_\perp = 0.014$ W/mK [54], we compute the transmission probability $\bar{\zeta}$ of the phonons through the CNT/SiO$_2$ interface. The phonon part of the thermal conductivity is

$$\kappa_\perp = \frac{1}{2\pi(T_{hot} - T_{cld})w_c} \int_0^\infty d\omega \cdot \hbar \omega \zeta_\omega \left( N_\omega^{hot} - N_\omega^{cld} \right)$$
$$\simeq \frac{\bar{\zeta}_\omega}{2\pi(T_{hot} - T_{cld})w_c} \int_0^\infty d\omega \cdot \hbar \omega \left( N_\omega^{hot} - N_\omega^{cld} \right), \quad (S11)$$

where the non-equilibrium distribution of phonons is approximated by the Bose-Einstein function with the effective temperatures $T_h$ and $T_c$ for the phonon subsystems in the "cold" (c) nanotube and the "hot" (h) substrate

$$N_\omega^{h(c)} = \frac{1}{\exp\left(\frac{\hbar \omega}{k_B T_{h(c)}}\right) - 1}. \quad (S12)$$

The above Eq. (S11) allows us to estimate the average transmission probability $\bar{\zeta}$ through the CNT/SiO$_2$ interface as

$$\bar{\zeta}_\omega = w_c \kappa_\perp \frac{2\pi(T_h - T_c)}{\int_0^\infty d\omega \cdot \hbar\omega (N_\omega^h - N_\omega^c)}$$
$$= w_c \kappa_\perp \frac{\hbar}{\pi k_B^2} \frac{12}{T_h + T_c},$$
(S13)

where $w_c$ is the effective contact width, and

$$\frac{1}{\hbar}\int_0^\infty d(\hbar\omega) \cdot \hbar\omega (N_\omega^{hot} - N_\omega^{cld}) = \frac{\pi^2 k_B^2}{6\hbar}(T_h - T_c)(T_h + T_c).$$
(S14)

Then using, e.g., $T_h + T_c = 85\,\text{K}$, $\kappa_\perp = 0.014\,\text{W}/(\text{m}\cdot\text{K})$, and the CNT/SiO2 contact width $w_c = 0.3\,\text{nm}$, we obtain

$$\bar{\zeta}_\omega = w_c \kappa_\perp \frac{\hbar}{\pi k_B^2} \frac{12}{T_h + T_c} \simeq 0.1,$$
(S15)

which agrees with the estimate (S10). The above Eqs. (S10) and (S15) suggest that only a small fraction of phonons (about 10%), propagating in the electrically active region, penetrate from the CNT into the SiO2 substrate, while most of them propagate along the CNT. However, outside the active C region, the scenario of phonon transport becomes quite different. One the one hand, abundance of the electric charge carriers in the L and R sections, serving as scattering centers for propagating phonons, causes the phonon transport along the CNT to be blocked. On the other hand, the phonon escape from the CNT into the substrate also stops, since the temperature gradient vanishes.

Besides, one can estimate the heat flux between the CNT and the SiO2 substrate using the Fourier law and the experimental results of Refs. [45-47]. Part of the thermal flux leaks from the CNT into the SiO2 substrate and is estimated to be

$$\vec{q} = -\kappa \nabla T$$
(S16)

where (in the SI units) $\vec{q}$ is the local heat flux density, $\text{Wm}^{-2}$, $\kappa$ is the material thermal conductivity, $\text{Wm}^{-1}\text{K}^{-1}$, $\nabla T$ is the temperature gradient, $\text{Km}^{-1}$. According to Refs. [45, 46], the thermal conductance $G$ of a stack involving the Au/Ti/graphene/SiO2 interfaces, is defined as

$$G = \frac{\delta \vec{q}}{\delta T} = -\kappa \frac{1}{\delta x}.$$
(S17)

For graphene multilayers with the number of graphene layers $1 \leq n \leq 10$ and in the temperature range of $50 \leq T \leq 500$ K they find $G \approx 25$ MW $\text{m}^{-2}\text{K}^{-1}$ irrespective of $n$ at room temperature, and that the heat flow across the metal/graphene/SiO$_2$ interface is limited by $G$ of the metal/graphene interface rather than by $G$ of the graphene/SiO$_2$ interface. Thus, the choice of metal contacts affects both electrical and thermal transport in graphene devices. Then the thermal conductance of the graphene/SiO$_2$ interface is

$$\kappa_{G/SiO_2} = -G\delta x = 0.025 \frac{W}{m \cdot K}, \tag{S18}$$

where we assumed the graphene/SiO$_2$ interface thickness $d_{G/SiO_2} = 1$ nm. Consequently, the thermal conductance $G$ of the Au/Ti/graphene/SiO$_2$ interface is

$$\kappa_{Ti/G/SiO_2} = -G\delta x = 0.006 \frac{W}{m \cdot K}. \tag{S19}$$

Using the Fourier law (S9), we compare the thermal flux along the CNT relatively to the thermal flux between the CNT and substrate. One estimates the thermal flux density between the CNT and the SiO$_2$ substrate as

$$Q_\perp = -\kappa_\perp \frac{T_h - T_c}{d_{G/SiO_2}} = 6.3 \times 10^{-8} \, W, \tag{S20}$$

which gives the ratio

$$\frac{Q_\perp}{Q_\parallel} = \frac{6.3 \times 10^{-8} \, W}{3.5 \times 10^{-6} \, W} \simeq 0.02, \tag{S21}$$

where for $Q_\parallel$ we used the estimation (S9). This means that only ~2% or less of the total heat flux is directed between the CNT and the SiO$_2$ substrate. This ratio is much larger for graphene stripes, whose contact area with the SiO$_2$ substrate is considerably larger. Therefore, for graphene stripes, which are overlaid on the SiO$_2$ substrate, the ratio $Q_\perp / Q_\parallel$ achieves a few percent.

## S4. Phonon drag

While the electric current passes the active region, it is carried by electrons on the left, and by holes on the right, propagating in opposite directions (see Fig. S2). If the current direction is positive, the electrons and holes are pushed away from the active region toward the metal electrodes; thus they suck the thermal energy from the active region, thereby transferring it further via the metal electrodes into the area outside of the contact. By feeding the electric current, e.g., $I = 0.01$ mA, one creates

$$\left(\frac{dn_e}{dt}\right)_{e-p} = \frac{1}{e} I = 6.2 \times 10^{13} \frac{1}{s} \tag{S22}$$

of electron-hole pairs per second. In Eq. (S22), the number of electrons created in C is $n_e = \sum_{\mathbf{p}} f_{\mathbf{p}}$, where $f_{\mathbf{p}}$ is the electron distribution function, determined by the Boltzmann equation. During their propagation, the electrons and holes transmit part of their energy to the phonons generated in the course of the electron-phonon collisions. The excited phonons propagate toward the outside area, and are eventually re-adsorbed back by the electrons and holes. However, according to Eqs. (S10), (S15), a smaller part (~10% of the total number) of propagating phonons escape from the CNT into the substrate, causing the heat flux leakage.

We assume that the leading mechanism responsible for creation of the electron and hole excitations in the active region C is determined by the electron-phonon collisions. The process of extraction of the charge carriers causes an energy drain leading to a decrease of the electron subsystem temperature $T_{el}$. Such extraction of the electron and hole excitations from the active region toward the metal electrodes is described by the particle balance equation

$$\sum_{\mathbf{p}} \left[ \left( \frac{\partial f_{\mathbf{p}}}{\partial t} \right)_{extr} - L_{e-p}(\mathbf{p}, f_{\mathbf{p}}) \right] = 0 , \qquad (S23)$$

where the first term in square brackets describes extraction of the quasiparticle electron and hole excitations from the active C region by electric current, while the second term, $L_{e-p}$, is the electron-phonon collision integral in the deformation potential approximation [36, 55-59]

$$\begin{aligned} L_{e-p}(\mathbf{k}, f_{\mathbf{k}}) = \sum_{\mathbf{q}} \frac{2|C_q|^2}{\Omega \hbar^2 |\varepsilon(\omega_q, \mathbf{q})|^2} \\ \times \{ [f_{\mathbf{k}}(1-f_{\mathbf{k+q}})N_{\mathbf{q}} - f_{\mathbf{k+q}}(1-f_{\mathbf{k}})(1+N_{\mathbf{q}})] \delta(\varepsilon_{k+q} - \varepsilon_k - \omega_q) \\ + [f_{\mathbf{k}}(1-f_{\mathbf{k-q}})(1+N_{\mathbf{q}}) - f_{\mathbf{k-q}}(1-f_{\mathbf{k}})N_{\mathbf{q}}] \delta(\varepsilon_k - \varepsilon_{k-q} - \omega_q) \}. \end{aligned} \qquad (S24)$$

The above Eqs. (S23), (S24) are complemented by the energy balance equations taking into account that the *e*- and *h*-excitations are created in the course of the electron-phonon collisions. The phonon energy loss rate per unit volume is found using the relation [60]

$$P = \frac{1}{\Omega} \sum_q \hbar \omega_q \frac{dN_q}{dt} + P_{esc}, \qquad (S25)$$

where $\Omega$ is the normalizing volume, and $P_{esc}$ is the heat leakage owing to the phonon escape into the substrate. Assuming that the leading contribution to $dN_q/dt$ in Eq. (S25) comes from the phonon-electron collisions [56, 61], one writes

$$\begin{aligned} \frac{dN_q}{dt} \simeq \frac{2|C_q|^2}{\Omega \hbar^2 |\varepsilon(\omega_q, \mathbf{q})|^2} \\ \times \sum_{\mathbf{k}} \{ [f_{\mathbf{k}}(1-f_{\mathbf{k+q}})N_{\mathbf{q}} - f_{\mathbf{k+q}}(1-f_{\mathbf{k}})(1+N_{\mathbf{q}})] \delta(\varepsilon_{k+q} - \varepsilon_k - \omega_q) \\ + [f_{\mathbf{k}}(1-f_{\mathbf{k-q}})(1+N_{\mathbf{q}}) - f_{\mathbf{k-q}}(1-f_{\mathbf{k}})N_{\mathbf{q}}] \delta(\varepsilon_k - \varepsilon_{k-q} - \omega_q) \}. \end{aligned} \qquad (S26)$$

Furthermore, for calculating $P$, we assume that the electron-electron collisions are sufficiently frequent in order to provide the use of the electron temperature approximation for the electrons. Using the above expressions (S25), (S26), one finds

$$P = \sum_q \frac{2|C_q|^2}{(\Omega\hbar)^2 |\varepsilon(\omega_q,\mathbf{q})|^2} \hbar\omega_q \left[N_q - N_q^{(c)}\right] \qquad (S27)$$
$$\times \sum_k \left[f_k^{(c)} - f_{k+q}^{(c)}\right] \delta(\varepsilon_{k+q} - \varepsilon_k - \omega_q),$$

where $N_q$ is the distribution function of "hot" phonons, $N_q^{(c)}$ is the distribution function of "cold" phonons, and $f_k^{(c)}$ is the distribution function of "cold" electrons. Next, we replace the summation in Eq. (S27) by integration as

$$\sum_k \left[f_k^{(c)} - f_{k+q}^{(c)}\right] \delta(\varepsilon_{k+q} - \varepsilon_k - \omega_q)$$
$$\to \int d\varepsilon_k g(\varepsilon_k) \left[f_{\varepsilon_k}^{(c)} - f_{\varepsilon_k+\omega_q}^{(c)}\right] \delta\left(v_F q + \frac{\hbar}{2m} q^2 - \omega_q\right), \qquad (S28)$$

where the finite electron mass $m$ arises due to presence of electron subbands (in the CNT, typically $m = 0.1 - 0.5 m_e$) and $g(\varepsilon)$ is the 1D electron density of states in the carbon nanotube [62]

$$g(\varepsilon) = \frac{\sqrt{3}}{\pi^2} \frac{1}{|V_{pp\pi}|} \frac{d}{r} \sum_{m=1}^{N} \mathrm{Re}\left(\frac{|\varepsilon|}{\sqrt{\varepsilon^2 - \varepsilon_m^2}}\right), \qquad (S29)$$

where $d$ is the carbon-carbon bond distance ($a = d\sqrt{3}$) and $r$ is the nanotube radius ($|R| = 2\pi r$). We use

$$\varepsilon_{k+q} - \varepsilon_k = \frac{\hbar^2 (k_F + q)^2}{2m} - \frac{\hbar^2 k_F^2}{2m}$$
$$= \frac{\hbar^2 k_F q}{m} + \frac{q^2 \hbar^2}{2m} \simeq v_F \hbar q + \frac{\hbar^2}{2m} q^2, \qquad (S30)$$

where $v_F = \hbar k_F / m$, which gives

$$\int \left[f_\varepsilon^{(c)} - f_{\varepsilon+\omega}^{(c)}\right] d\varepsilon \simeq T \log\left(\frac{2 e^{\frac{\omega_q}{T_c}}}{e^{\frac{\omega_q}{T_c}} + 1}\right). \qquad (S31)$$

Using the above formulas (S28)-(S30), Eq. (S27) is rewritten as

$$P = N(0) T_c \frac{2|C_0|^2}{(\Omega\hbar)^2 |\varepsilon_{ph}^0|^2} \sum_q \hbar\omega_q \left[\frac{1}{e^{\frac{\omega_q}{T_h}} - 1} - \frac{1}{e^{\frac{\omega_q}{T_c}} - 1}\right]$$
$$\times \log\left(\frac{2 e^{\frac{\omega_q}{T_c}}}{e^{\frac{\omega_q}{T_c}} + 1}\right) \delta\left(v_F q + \frac{\hbar}{2m} q^2 - \omega_q\right). \qquad (S32)$$

The last Eq. (S32) allows one to compute the loss rate of thermal energy transmitted from "hotter" phonons with the temperature $T_h$ to the colder electrons characterized by an effective temperature $T_c$ due to their expulsion from the active region caused by the flowing electric current. The integral (S32) is simplified to the form

$$P = N(0)T_c \frac{2|C_0|^2}{(\Omega\hbar)^2 |\varepsilon_{ph}^0|^2} \int_0^{\omega_D} d\omega_q F(\omega_q) \hbar\omega_q \left[ \frac{1}{e^{\frac{\omega_q}{T_h}} - 1} - \frac{1}{e^{\frac{\omega_q}{T_c}} - 1} \right]$$

$$\times \log\left( \frac{2e^{\frac{\omega_q}{T_c}}}{e^{\frac{\omega_q}{T_c}} + 1} \right) \delta\left( v_F q + \frac{\hbar}{2m} q^2 - \omega_q \right) \quad \text{(S33)}$$

$$\simeq N(0)T_c^2 \frac{2|C_0|^2}{(\Omega\hbar)^2 |\varepsilon_{ph}^0|^2} F(\omega_q^{(1)}) \Phi(T_h, T_c)$$

where $\hbar\omega_q^{(1)} = 2p_F s$, $s$ is the sound velocity and $F(\omega_q)$ is the phonon density of states, and factor

$$\Phi(\lambda) = x \left[ \frac{1}{e^{x/\lambda} - 1} - \frac{1}{e^x - 1} \right] \log\left( \frac{2e^x}{e^x + 1} \right), \quad \text{(S34)}$$

and

$$x = \frac{2mv_F s}{k_B T_c}, \quad \lambda = \frac{T_h}{T_c}, \quad \frac{2mv_F s}{k_B T_c} \frac{T_c}{T_h} = \frac{x}{\lambda}, \quad \text{(S35)}$$

where $v_F = 2 \times 10^6$ m/s is the Fermi velocity and $s = 2 \times 10^6$ cm/s is the sound velocity in the CNT. In the above equations, the screening effects were introduced by dividing the matrix elements $C_0$ by the dielectric function of graphene $\varepsilon_{ph}^0$. Taking into account that the matrix elements in graphene are determined by the change in the overlap between the orbitals surrounding different atoms and not by a Coulomb potential, we set $\varepsilon_{ph}^0 = 1$. Depending on the geometry and on the ratio $T_h/T_{cld} = 1-100$, $\Phi(\lambda, x)$ achieves values of $10-10^3$, which gives an estimate $P \simeq 0.1-10$ nW.

The above estimation was obtained assuming that the "hot" temperature $T_h = 77$ K and the "cold" temperature $T_c = 0.8-8$ K. Furthermore, we assumed that phonon confinement does not influence the phonon-electron scattering rates. Here we used the same parameters [35, 59, 63] as listed above in Eq. (S4) and

$$g(0) \simeq \frac{\sqrt{3}}{\pi^2} \frac{1}{|V_{pp\pi}|} \frac{2d_{cc}}{d_{CNT}} = 6.6454 \times 10^{-3} \frac{1}{\text{eV}} \quad \text{(S36)}$$

where $d_{cc} = 0.142$ nm is the spacing of the carbon atom bonds ($a = d_{cc}\sqrt{3}$), $V_{pp\pi} \approx 2.5$ eV is the nearest-neighbor pp interaction.

Inside the active C-region, the "hot" temperature $T_{ph}$ of the phonon subsystem, as well as the "cold" temperature $T_{el}$ of the electron subsystem, are computed using the above particle and energy balance equations (S23) and (S25). Typical result of the numeric calculations is shown in Fig. S1a. In this scenario, the energy deficit of the electron subsystem is compensated by the energy obtained from the phonon subsystem in the course of the electron-phonon collisions. Furthermore, the electron-phonon scattering cause generation of a sufficient number of the new electron-hole pairs to maintain the electric current, which is accompanied by the outflow of charge carries from the active region. Using the energy and quasiparticle balance equations, we obtain two conditions determining the local effective temperature of electrons $T_{el}$ and phonons $T_{ph}$ inside the active region.

Taking into account that the mean free path $l_{e(h)}$ of the electrons (*e*) and holes (*h*) in the active region is about $l_{e(h)} = 1.3$ μm, we conclude that the change carriers, while propagating in the active region, transfer a smaller part (<7%) of their energy to phonons. The larger fraction (>93%) of the electrons and holes reaches the area outside the active region and goes to the metal electrodes. The phonon mean free path $l_{p-p}$ in the active region is determined mostly by the phonon-phonon scattering, which is relatively weak, therefore $l_{p-p}$ exceeds the length of the active region, i.e., $l_{p-p} > 1$ μm, meaning that phonons practically do not collide with electrons and holes inside the neutral active region in the C section. Instead, a larger fraction (>90%) of phonons propagates along the nanotube, while their smaller part (<10%) goes into the SiO$_2$ substrate. Upon reaching the gated sections of the nanotube, the phonons collide there with the charge carriers, and their mean free path becomes very short ($l_{p-e} < 0.3$ μm). For such reasons, the phonons quickly transfer their energy to the charge carriers which tunnel from the CNT into the metal electrodes thereafter.

## S5. Quantum capacitance

In the experiments with the carbon nanotube field effect transistors (FET), the effect of the capacitance [64-66] is evaluated as follows. Knowing the length of a device, one obtains a gate efficiency (typically α₁ = 0.1-1%), which allows one to determine the quantum capacitance of the nanotube $C_q$ using the equation

$$C_q = C_{ge} \frac{1-\alpha}{\alpha}, \tag{S37}$$

where $C_{ge} \approx 2\pi\varepsilon L / \ln(4h/d)$ is the geometry capacitance of the nanotube, *h* is the thickness of the dielectric SiO$_2$ between the doped Si substrate and the nanotube, $\varepsilon = 3.9\varepsilon_0$ for SiO$_2$, and *d* is the diameter of the nanotube. Typical geometry capacitance of the CNT device is ~10 aF, and the quantum capacitance is $C_q$~1000-2000 aF.

The total voltage change is the sum of these two contributions. Therefore, the total effect is *as if* there are two capacitances in series: The conventional geometry-related capacitance $C_{ge}$ (as calculated by the Gauss law), and the "quantum capacitance" $C_q$ related to the density of states [64-66]. The latter is

$$C_q = e^2 N(0), \qquad (S38)$$

where $N(0)$ is the electron density of states at the Fermi level (S36). Eqs. (S37), (S38) suggest that neither the geometrical capacitance $C_{ge}$ nor the quantum capacitance $C_q$ depend on temperature, since the electron density of states $N(0)$ is temperature-independent. A thorough derivation of formula for $C_q$ shows that the temperature dependence basically is pronounced as a change of the width $\Gamma(T)$ of sharp peaks, occurring in the dependence of electric differential conductance $G_e(V_{SD})$ versus the source-drain bias voltage $V_{SD}$. A most essential contribution to $\Gamma(T)$ originates from the inelastic processes of the electron-phonon scattering, which are responsible for the energy dissipation in the system. An inelastic effect of the electron-phonon scattering is accounted by replacing in Eq. (S29) the electron energy as $\varepsilon \to \varepsilon + i\Gamma(T)$, which results in the temperature dependence of the electron spectral singularities, and can be observed experimentally by measuring the electric differential conductance $G_e(V_{SD})$. Therefore, the temperature dependence of the electron peak width $\Gamma(T)$, after a proper calibration, can be used as a meter for monitoring the effective electron temperature $T_{el}$. This approach is used in our work to determine the effect of intrinsic cooling in our CNT FET setup.

**References**

1. Bell, L. E. *Science* **2008,** 321, (5895), 1457-1461.
2. DiSalvo, F. J. *Science* **1999,** 285, (5428), 703-706.
3. Berber, S.; Kwon, Y. K.; Tomanek, D. *Phys Rev Lett* **2000,** 84, (20), 4613-4616.
4. Dubi, Y.; Di Ventra, M. *Rev Mod Phys* **2011,** 83, (1), 131-155.
5. Finch, C. M.; Garcia-Suarez, V. M.; Lambert, C. J. *Phys Rev B* **2009,** 79, (3).
6. Pop, E.; Mann, D. A.; Goodson, K. E.; Dai, H. J. *J Appl Phys* **2007,** 101, (9).
7. Reddy, P.; Jang, S. Y.; Segalman, R. A.; Majumdar, A. *Science* **2007,** 315, (5818), 1568-1571.
8. Shafranjuk, S. E. *Epl-Europhys Lett* **2009,** 87, (5).
9. Small, J. P.; Perez, K. M.; Kim, P. *Phys Rev Lett* **2003,** 91, (25).
10. Heremans, J. P.; Dresselhaus, M. S.; Bell, L. E.; Morelli, D. T. *Nat Nanotechnol* **2013,** 8, (7), 471-473.
11. Zhao, L. D.; Lo, S. H.; Zhang, Y. S.; Sun, H.; Tan, G. J.; Uher, C.; Wolverton, C.; Dravid, V. P.; Kanatzidis, M. G. *Nature* **2014,** 508, (7496), 373-+.
12. Snyder, G. J.; Toberer, E. S. *Nat Mater* **2008,** 7, (2), 105-114.
13. Shafranjuk, S. E. *Eur Phys J B* **2014,** 87, (4).
14. Shafranjuk, S. E. *Epl-Europhys Lett* **2015,** 109, (6).
15. Mayle, S.; Gupta, T.; Davis, S.; Chandrasekhar, V.; Shafraniuk, S. *J Appl Phys* **2015,** 117, (19).
16. Shafraniuk, S., *Graphene: Fundamentals, Devices and Applications*. Pan Stanford: 2015; p 634.
17. Rowe, D. M., *Thermoelectrics handbook : macro to nano*. CRC/Taylor & Francis: Boca Raton, 2006.
18. Avouris, P.; Freitag, M.; Perebeinos, V. *Nature Photonics* **2008,** 2, (6), 341-350.
19. Rinzan, M.; Jenkins, G.; Drew, H. D.; Shafranjuk, S.; Barbara, P. *Nano Lett* **2012,** 12, (6), 3097-

3100.
20. Takahashi, Y.; Nagase, M.; Namatsu, H.; Kurihara, K.; Iwdate, K.; Nakajima, K.; Horiguchi, S.; Murase, K.; Tabe, M. *Electron Lett* **1995,** 31, (2), 136-137.
21. Xiong, Z. W.; Wang, X. M.; Yan, D. W.; Wu, W. D.; Peng, L. P.; Li, W. H.; Zhao, Y.; Wang, X. M.; An, X. Y.; Xiao, T. T.; Zhan, Z. Q.; Wang, Z.; Chen, X. R. *Nanoscale* **2014,** 6, (22), 13876-13881.
22. Zheng, H. S.; Asbahi, M.; Mukherjee, S.; Mathai, C. J.; Gangopadhyay, K.; Yang, J. K. W.; Gangopadhyay, S. *Nanotechnology* **2015,** 26, (35).
23. Bhadrachalam, P.; Subramanian, R.; Ray, V.; Ma, L. C.; Wang, W. C.; Kim, J.; Cho, K.; Koh, S. J. *Nature Communications* **2014,** 5.
24. Shafraniuk, S. E., Nanosensors of External Fields. In *Encyclopedia of Nanoscience and Nanotechnology*, Nalwa, H. S., Ed. American Scientific Publishers: 2011; Vol. 18, pp 413-454.
25. Nakai, Y.; Honda, K.; Yanagi, K.; Kataura, H.; Kato, T.; Yamamoto, T.; Maniwa, Y. *Applied Physics Express* **2014,** 7, (2).
26. Zuev, Y. M.; Chang, W.; Kim, P. *Phys Rev Lett* **2009,** 102, (9).
27. Gerritsen, J. W.; Shafranjuk, S. E.; Boon, E. J. G.; Schmid, G.; vanKempen, H. *Europhys Lett* **1996,** 33, (4), 279-284.
28. Y. Yang, G. F., S. E. Shafranjuk, T. M. Klapwijk, B. K. Cooper, R. M. Lewis, C. J. Lobb, and P. Barbara. *Nano Lett* **2015**.
29. Yang, Y. F.; Fedorov, G.; Zhang, J.; Tselev, A.; Shafranjuk, S.; Barbara, P. *Supercond Sci Tech* **2012,** 25, (12).
30. Hicks, L. D.; Dresselhaus, M. S. *Phys Rev B* **1993,** 47, (24), 16631-16634.
31. Hicks, L. D.; Dresselhaus, M. S. *Phys Rev B* **1993,** 47, (19), 12727-12731.
32. Hicks, L. D.; Dresselhaus, M. S. *Semiconductor Heterostructures for Photonic and Electronic Applications* **1993,** 281, 821-826.
33. Lin, Y. M.; Sun, X. Z.; Dresselhaus, M. S. *Phys Rev B* **2000,** 62, (7), 4610-4623.
34. Zhang, Z. B.; Sun, X. Z.; Dresselhaus, M. S.; Ying, J. Y.; Heremans, J. *Phys Rev B* **2000,** 61, (7), 4850-4861.
35. Adam, S.; Hwang, E. H.; Das Sarma, S. *Physica E* **2008,** 40, (5), 1022-1025.
36. Hwang, E. H.; Das Sarma, S. *Phys Rev B* **2008,** 77, (11).
37. Hwang, E. H.; Das Sarma, S. *Phys Rev B* **2008,** 77, (8).
38. Gautreau, P.; Chu, Y. B.; Ragab, T.; Basaran, C. *Computational Materials Science* **2015,** 103, 151-156.
39. Chu, Y. B.; Gautreau, P.; Basaran, C. *Appl Phys Lett* **2014,** 105, (11).
40. Gautreau, P.; Ragab, T.; Chu, Y. B.; Basaran, C. *J Appl Phys* **2014,** 115, (24).
41. Gautreau, P.; Ragab, T.; Basaran, C. *Carbon* **2013,** 57, 59-64.
42. Gautreau, P.; Ragab, T.; Basaran, C. *J Appl Phys* **2012,** 112, (10).
43. Lindsay, L.; Broido, D. A. *Phys Rev B* **2010,** 82, (20).
44. Zou, J.; Balandin, A. *J Appl Phys* **2001,** 89, (5), 2932-2938.
45. Koh, Y. K.; Bae, M. H.; Cahill, D. G.; Pop, E. *Nano Lett* **2010,** 10, (11), 4363-4368.
46. Cahill, D. G.; Braun, P. V.; Chen, G.; Clarke, D. R.; Fan, S. H.; Goodson, K. E.; Keblinski, P.; King, W. P.; Mahan, G. D.; Majumdar, A.; Maris, H. J.; Phillpot, S. R.; Pop, E.; Shi, L. *Applied Physics Reviews* **2014,** 1, (1).
47. Hsieh, W. P.; Lyons, A. S.; Pop, E.; Keblinski, P.; Cahill, D. G. *Phys Rev B* **2011,** 84, (18).
48. Yamamoto, T.; Konabe, S.; Shiomi, J.; Maruyama, S. *Applied Physics Express* **2009,** 2, (9).
49. Ramu, A. T.; Cassels, L. E.; Hackman, N. H.; Lu, H.; Zide, J. M. O.; Bowers, J. E. *J Appl Phys* **2011,** 109, (3).
50. Ramu, A. T.; Cassels, L. E.; Hackman, N. H.; Lu, H.; Zide, J. M. O.; Bowers, J. E. *J Appl Phys* **2010,** 107, (8).
51. Parrott, J. E. *Revue Internationale Des Hautes Temperatures Et Des Refractaires* **1979,** 16, (4), 393-403.
52. Ando, T. *J Phys Soc Jpn* **2005,** 74, (3), 777-817.

53. Marconnet, A. M.; Panzer, M. A.; Goodson, K. E. *Rev Mod Phys* **2013,** 85, (3), 1295-1326.
54. Chen, L.; Kumar, S. *Journal of Heat Transfer-Transactions of the Asme* **2014,** 136, (5).
55. Ziman, J. M. *Philos Mag* **1965,** 11, (110), 438-&.
56. Ziman, J. M., *The Theory of Transport Phenomena in Solids*. Clarendon Press: Oxford, 1960.
57. Hwang, E. H.; Adam, S.; Das Sarma, S. *Phys Rev Lett* **2007,** 98, (18).
58. Hwang, E. H.; Das Sarma, S. *Phys Rev B* **2007,** 75, (20).
59. Hwang, E. H.; Das Sarma, S. *Phys Rev B* **2008,** 77, (19).
60. Bockelmann, U.; Bastard, G. *Phys Rev B* **1990,** 42, (14), 8947-8951.
61. Ziman, J. M. *Philos Mag* **1956,** 1, (2), 191-198.
62. Mintmire, J. W.; White, C. T. *Phys Rev Lett* **1998,** 81, (12), 2506-2509.
63. Hwang, E. H.; Das Sarma, S. *Phys Rev Lett* **2008,** 101, (15).
64. Ilani, S.; Donev, L. A. K.; Kindermann, M.; McEuen, P. L. *Nat Phys* **2006,** 2, (10), 687-691.
65. Parkash, V.; Goel, A. K. *Nanoscale Research Letters* **2010,** 5, (9), 1424-1430.
66. Parkash, V.; Goel, A. K. *2010 International Conference on Microelectronics* **2010**, 244-247.